\documentclass[10pt, conference, letterpaper]{IEEEtran}

\ifCLASSINFOpdf
   \usepackage[pdftex]{graphicx}
   \graphicspath{{../pdf/}{../jpeg/}}
   \DeclareGraphicsExtensions{.pdf,.jpeg,.png}
\else
   \usepackage[dvips]{graphicx}
   \graphicspath{{../eps/}}
   \DeclareGraphicsExtensions{.eps}
\fi

\usepackage{amsmath}
\usepackage{amsfonts}
\usepackage{marvosym}
\usepackage{makeidx}
\usepackage{mathrsfs}
\usepackage{url}
\usepackage{graphicx}
\usepackage{algorithm}
\usepackage{algorithmic}
\usepackage[english]{babel}
\usepackage{fullpage}
\usepackage{setspace}

\usepackage{graphicx}
\DeclareGraphicsExtensions{.jpg}
\usepackage{color}
\newcommand{\be}{\begin{equation}}
\newcommand{\ee}{\end{equation}}

\begin{document}

\title{Nonuniform probability modulation for reducing energy consumption of remote sensors}
\author{\IEEEauthorblockN{Jarek Duda}
\IEEEauthorblockA{Jagiellonian University, Golebia 24, 31-007 Krakow, Poland. Email: \emph{dudajar@gmail.com}}}
\maketitle

\begin{abstract}
One of the main goals of 5G wireless telecommunication technology is improving energy efficiency, especially of remote sensors which should be able for example to transmit on average 1bit/s for 10 years from a single AAA battery. There will be discussed using modulation with nonuniform probability distribution of symbols for improving energy efficiency of transmission at cost of reduced throughput. While the zero-signal (silence) has zero energy cost to emit, it can carry information if used alongside other symbols. If used more frequently than others, for example for majority of time slots or OFDM subcarriers, the number of bits transmitted per energy unit can be significantly increased. For example for hexagonal modulation and zero noise, this amount of bits per energy unit can be doubled by reducing throughput 2.7 times, thanks to using the zero-signal with probability $\approx$ 0.84. There will be discussed models and methods for such nonuniform probability modulations (NPM).
\end{abstract}

\IEEEpeerreviewmaketitle
\section{Introduction}
The currently being developed 5th generation mobile network (5G) has many ambitious goals, like 10Gbps peak data rates, 1ms latency and ultra-reliability. Another high  priority is reducing energy consumption, especially to improve battery life of mobile and IoT devices. This goal is crucial for expected omnipresent fleet of remote sensors, monitoring all aspects of our world. Such sensor should be compact, inexpensive and has battery life of order of 10 years, as battery replacement in many applications is economically infeasible. Hence, this is an asymmetric task: the main priority is to reduce energy requirements of the sender, tolerating increased cost at the receiver side. A crucial part of the cost of sending information to the base station is establishing connection - their number and so energy cost can be reduced by buffering information, or can be nearly eliminated if the sensor just transmits the information in time periods precisely scheduled with the base station.

We will discuss approach for reducing energy need of the actual transmission of such buffered data, preferably compressed earlier to reduce its size. The information is encoded in a sequence of symbols as points from a chosen constellation: a discrete set of points in complex (I-Q) plane. This sequence of symbols can be used for time sequence of impulses, or as coefficients for usually orthogonal family of functions (subcarriers), for example in OFDM. 

These constellations are often QAM lattices of size up to 64 in LTE. There is assumed uniform probability modulation (UPM) - that every symbol is used with the same frequency. Generally, a stream of symbols having ${p_s}_s$ probability distribution $(\sum_s p_s =1)$ contains asymptotically Shannon entropy $h=\sum_{s=1}^m p_s \lg(1/p_s)$ bits/symbol ($\lg\equiv \log_2$), where $m$ is the size of alphabet. Entropy is indeed maximized for uniform probability distribution $p_s = 1/m$, obtaining $h=\lg(m)$ bits/symbol, which is $\lg(64)=6$ bits/symbol for QAM64.

However, this natural choice of using uniform probability distribution is not always the optimal one. For example when the channel has constraints, like forbidding two successive ones ('11') in Fibonacci coding, then choosing $\Pr(x_{t+1}=0 | x_t=0) = \Pr(x_{t+1}=1 | x_t=0) = 1/2$ is not the optimal way. Instead, we should more often choose '0' symbol, optimally with $\varphi =(\sqrt{5}-1)/2$ probability, as this symbol allows to produce more entropy (information) in the successive step. For a general constraints the optimal probabilities can be found by using Maximal Entropy Random Walk~\cite{d2007}.

Another example of usefulness of nonuniform probability distribution among symbols used for communication are various steganography-watermarking problems, where we want encoding sequence to resemble some common data, for example picture resembling QR codes. Surprisingly, generalization of the Kuznetsov-Tsybakov problem allows such encoding without decoder knowing the used probability distributions (e.g. picture to resemble)~(\cite{KT1, KT}). However, this lack of knowledge makes encoding more expensive. \\

In this paper we will focus on a more basic reason to use nonuniform probability distribution among symbols: that the cost of using various symbols does not have to be the same. Assume $E_s$ is the cost of using symbol $s$, then entropy for a fixed average energy ($E=\sum_s p_s E_s$) is maximized for Boltzmann probability distribution among symbols $\left(\Pr(s)\propto e^{-\beta E_s} \right)$. For example in Morse code $dash$ lasts much longer than $dot$, what comes with higher time and energy cost. Designing a coding with more frequent use of dot ($\Pr(dot) > \Pr(dash)$) would allow to lower average cost per bit. Anther example of nonuniform cost is sending symbol '1' as electric current through a wire, symbol '0' as lack of this current - symbol '1' is more energy costly, hence should be used less frequently. \\

We will focus here on application for wireless communication modulation, where the cost related to emitting a symbol is usually assumed to be proportional to square of its amplitude: $E_x \propto |x|^2$, hence we could improve energy efficiency by more frequent use of low amplitude symbols.

Basic theoretical considerations will be reminded, then used to analyze potential improvements especially for the situation of modulation for wireless technology: to reduce required energy per bit, especially for the purpose of improving battery life of remote sensors. The average amount of bits/symbol (entropy) is maximized for uniform probability distribution (UPM), hence using nonuniform distribution (NPM) means that more symbols are required to write the same message, so the tradeoff of improving energy efficiency (bits per energy unit) is also reducing throughput (bits per symbol).

The use of nonuniform probability distribution of symbols requires a more complex coding scheme, especially from the perspective of error correction (channel coding). Entropy coders allow to work with kind of reversed task: encode a sequence of symbols having some assumed probability distribution into a bit sequence. Switching its encoder and decoder, we can encode a message (a bit sequence) into a sequence of symbols having some chosen probability distribution. Due to low cost, a natural approach would be using a prefix code here, for example $0\to a,\ 10\to b,\ 11\to c$. However, it approximates probabilities with powers of $1/2$ and cannot use probabilities $1/2 < p < 1$, which turn out crucial in the discussed situations. Additionally, its error correction would require some additional protection layer. Hence, a more appropriate recent entropy coding will be discussed for this purpose: tANS coding~(\cite{ANS,ANS1}). While having cost similar to prefix codes (finite state automaton, no multiplication), it operates on nearly accurate probabilities, including $1/2 < p < 1$. Additionally, its processing has an internal state, which can be exploited like the state of convolutional codes~\cite{conv} for error correction purpose - thanks of it encoder does not need to apply another coding layer, saving energy required for this purpose.

\section{Capacity and energy efficiency of nonuniform probability modulation (NPM)}
In this section there will be first reminded why Boltzmann distribution is the optimal choice from the perspective of energy efficiency, then three modulations will be analyzed, first without then with noise.

For better intuition, Shannon entropy is measured in bits: $h=\sum_{s=1}^m p_s \lg(1/p_s)$ bits/symbol ($\lg\equiv \log_2$).

\subsection{Probability distribution maximizing entropy}
Assume $E_s$ is the cost (energy) of using symbol $s$. We want to choose the optimal probability distribution $\{p_s\}_s$ for some fixed average energy $E$:
\be \sum_s p_s E_s = E \qquad\qquad\qquad \sum_s p_s =1 \ee
such that Shannon entropy is maximized: $h \ln(2) =- \sum_s p_s \ln(p_s)$.

Using the Lagrange multiplier method for $\lambda$ and $\beta$ parameters:
$$ L= - \sum_s p_s \ln p_s + \lambda \left( \sum_s p_s - 1 \right) + \beta \left(\sum_s p_s E_s - E\right) $$
$$0 = \frac{\partial L}{\partial p_s}=  -\ln(p_s) - 1 + \lambda + \beta E_s$$
\be p_s = \frac{e^{-\beta E_s}}{e^{1-\lambda}} = \frac{e^{-\beta E_s}}{Z} \ee
where $Z = e^{1-\lambda} = \sum_s e^{-\beta E_s}$ is the normalization factor (called partition function).

The parameter $\beta$ can be determined from average energy:
$$E=\frac{\sum_s E_s e^{-\beta E_s}}{\sum_s e^{-\beta E_s}}$$

As expected, Boltzmann distribution is the optimal way to choose probability distribution of symbols: $p_s \propto e^{-\beta E_s}$. The standard way of evaluating cost of a signal in wireless telecommunication is square of its amplitude: $E_s = |x|^2$. Hence for $x\in\mathbb{R}$ the optimal probability is Gaussian distribution with standard deviation $\sigma^2=E$:

$$\rho_G(x)=\frac{1}{\sqrt{2E\pi}} e^{\frac{-x^2}{2E}}$$
\be H_G :=-\int_{-\infty}^{\infty} \rho_G(x) \lg(\rho_G(x)) dx = \frac{1}{2}\lg(2\pi e E)  \ee
Let us compare it with uniform distribution, which is usually used in practical modulation schemes. Take a rectangular density function on some $[-a,a]$ range with height $\frac{1}{2a}$ to integrate to 1. Its average energy is $ E=\int_{-a}^a \frac{1}{2a} x^2 dx =\frac{a^2}{3}$, getting $a=\sqrt{3E}$ parameter for a chosen average energy $E$. Now
$$ H_u :=\int_{-a}^{a} \frac{1}{2a} \lg(2a) dx =\lg(2a)=\frac{1}{2}\lg(12E)$$
So the gain of using Gaussian distribution is
\be H_G - H_u =\frac{1}{2}\lg(\pi e /6) \approx 0.2546\ \textrm{bits}. \ee

There was used differential entropy (with integrals), which gets natural intuition when approximated with Riemann integration for some quantization step $q$:
$$H=-\int_{-\infty}^{\infty} \rho(x) \lg(\rho(x)) dx \approx -\sum_{k\in\mathbb{Z}} q\rho(kq) \lg(\rho(kq))=$$
$$=-\sum_{k\in\mathbb{Z}} q\rho(kq) \lg(q\rho(kq)) + \sum_{k\in\mathbb{Z}} q\rho(kq) \lg(q)$$
The left hand side term is the standard entropy for probability distribution of quantization with step $q$, the right hand side term is approximately $\lg(1/q)$. So entropy for quantized continuous probability distribution is approximately the differential entropy plus $\lg(1/q)$:
\be h_q \approx H + \lg(1/q) \ee

\begin{figure}[t!]
    \centering
        \includegraphics[width=8cm]{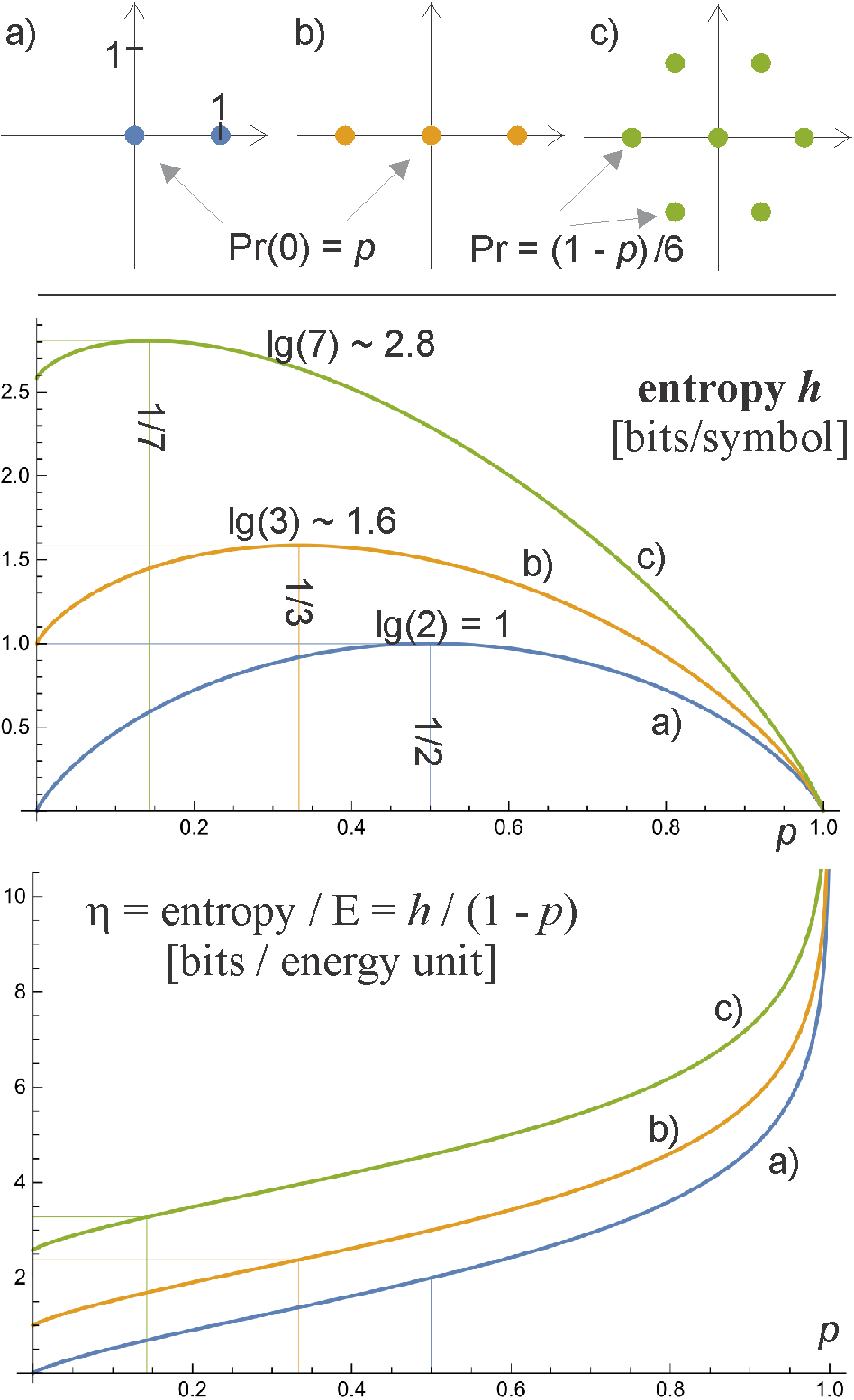}\
\begin{center}
        \caption{Top: three considered modulations (binary, ternary and hexagonal). Probability of using the zero signal is denoted by $p$, the remaining symbols have equal probability. Middle: dependence of entropy (bits/symbol) from the parameter $p$. It is maximized when all symbols are equally probable (UPM, marked). Bottom: energy efficiency (bits/energy unit) as this entropy divided by average amount of energy used per symbol - it tends to infinity when $p\to 1$, what means communicating by rarely distracting the silence of using the zero-signal. Energy efficiency is improved at cost of reducing throughput, which is proportional to entropy (assuming zero noise). For example for hexagonal modulation, UPM $(p=1/7)$ allows to transmit $\approx 3.275$ bits/energy unit. We can double it by using $p\approx 0.84$ frequency of the zero-signal, at cost of $\approx 2.7$ times lower entropy (throughput). Binary modulation, for example for wire communication, has even larger potential for improvements (e.q. to quadruple efficiency).}
   \label{entr}
\end{center}
\end{figure}

\subsection{Three modulations with zero-signal}
As discussed, the gain of using the optimal: Gaussian distribution in contrast to the standard uniform distribution, is $\approx 0.2546$ bits/symbol. Surprisingly, this is an absolute difference - it can bring an arbitrarily large relative difference for low original throughput: low $E$ and sparse quantization (large $q$).

So let us consider 3 basic examples of such modulation, visualized in the top of Fig. \ref{entr}:

a) \emph{binary}: $x\in \{0,1\}$

b) \emph{ternary} $x \in \{-1,0,1\}$

c) \emph{hexagonal} $x\in \{0\} \cup \{e^{i k \pi/3}: k=0,\ldots 5\}$

All of them contain zero-signal: which energy cost is zero (neglecting additional costs). This symbol represents information by using silence. Obviously, other symbols are also required - storing information by choosing moments (or subcarriers) to break this silence. Hexagonal modulation is appropriate for wireless communication. Binary and ternary are less effective, but they can be useful for communication by wire.

For all three cases denote $\Pr(0)=p$ as the probability of using zero signal. The remaining signals have all energy cost $E_x=|x|^2=1$, hence we can assume uniform probability distribution among them (correspondingly: $1-p$, $(1-p)/2$, $(1-p)/6$). The average energy in all three cases is
\be E=p\cdot 0 + (1-p)\cdot 1 = 1-p.\ee
The middle plot of Fig. \ref{entr} shows entropy dependence of $p$ for all three cases: average number of bits/symbol. It is maximized for UPM: $p=1/2,\ 1/3$, or $1/7$ correspondingly (marked). However, if we divide entropy by average energy cost of a symbol, getting average bits/energy unit, this energy efficiency $\eta=h/E$ grows to infinity for $p\to 1$, at cost of reduced entropy (throughput).

\subsection{Adding white Gaussian noise}
In real-life scenarios we also need to take noise into consideration: sender adds some redundancy to the transmitted message, then receiver applies forward error correction to repair eventual errors thanks to using this redundancy. The Shannon noisy-channel coding theorem~\cite{shannon} says that capacity of a channel is:
\be C =\max_{p_X} I(X;Y) = \max_{p_X} h(Y) - h(Y|X) \ee
Without the maximization, mutual information $I(X;Y)$ determines how many bits/symbol can on average be sent through the channel (including error correction), assuming the sender uses $p_X$ probability distribution among symbols. Capacity $C$ uses probability distribution miximizing the throughput, in analogy to maximizing entropy by UPM in the previously considered noise-free case. In contrast, we will focus on priority of optimizing energy efficiency here:
\be \eta = \frac{I(X;Y)}{E}=\frac{I(X;Y)}{\sum_x \Pr(x)E_x}\quad \textrm{bits per energy unit}\ee

\begin{figure}[t!]
    \centering
        \includegraphics[width=8cm]{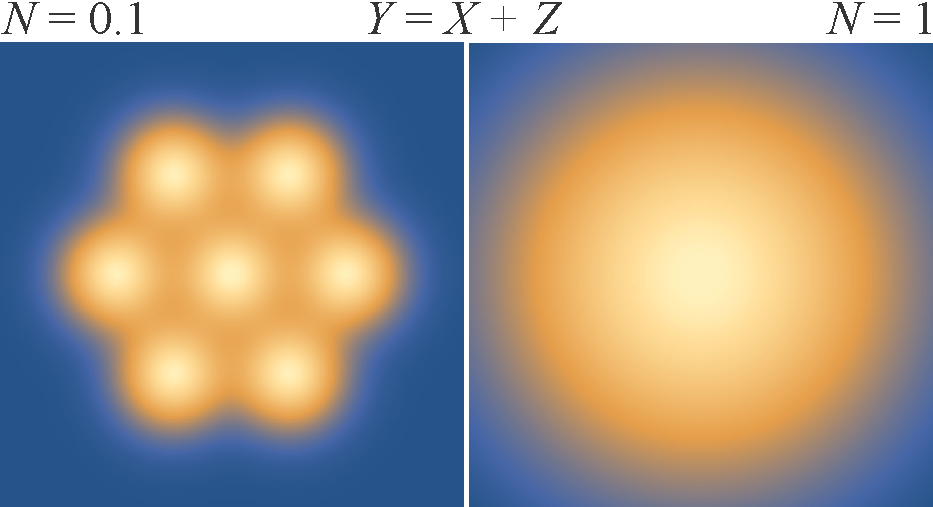}\
\begin{center}
        \caption{Density of $Y=X+Z$ for hexagonal modulation and $N=0.1$ (left) or $N=1$ (right), assuming noise $Z$ from two-dimensional Gaussian distribution.}
   \label{dens}
\end{center}
\end{figure}

\begin{figure}[t!]
    \centering
        \includegraphics[width=8cm]{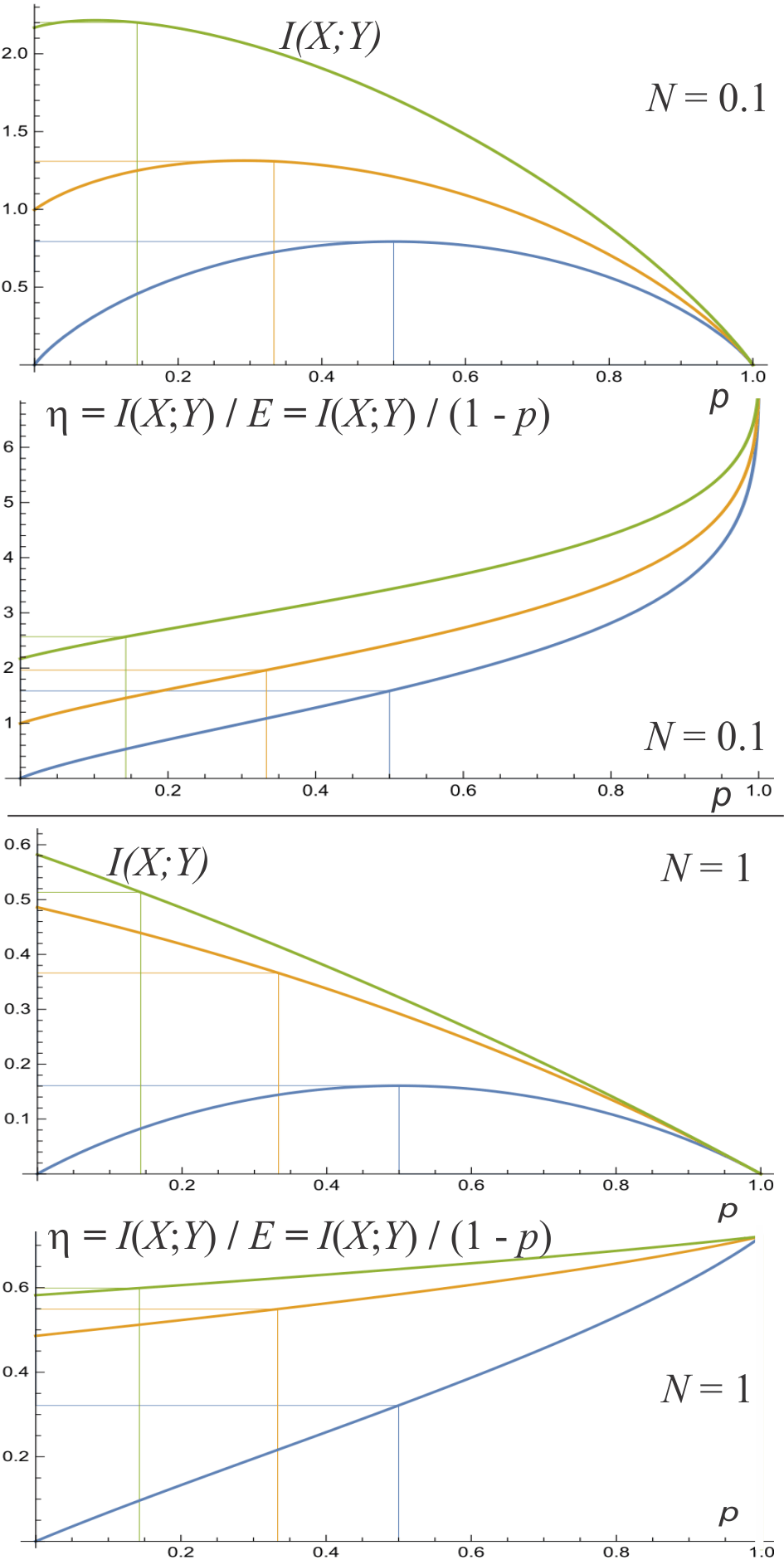}\
\begin{center}
        \caption{Four plots of correspondingly: entropy and efficiency for $N=0.1$, entropy and efficiency for $N=1$ for all three modulations.}
   \label{mutinf}
\end{center}
\end{figure}

In the case of modulation, there is usually assumed Gaussian noise:
\be Y=X+Z\qquad \textrm{where}\quad Z \sim \mathcal{N}(0,N) \ee
is from Gaussian distribution with $\sigma^2=N$ average energy of noise (variance). We will assume two-dimensional Gaussian noise in complex plane:
$$ \rho_Z(z)=\frac{1}{2\pi N} e^{-\frac{|z|^2}{2N}}$$

Figure \ref{dens} presents examples of $Y=X+Z$ density. After fixing some $X$ value, $Y=X+Z$ is Gaussian distribution around this value, hence $h(Y|X)=h(Z)$. To find capacity, we need to find probability distribution for $X$ to maximize $h(X+Y)$. For a fixed average energy $E$ of $X$, this entropy is maximized for $X\sim \mathcal{N}(0,E)$ Gaussian distribution, getting $Y=X+Y\sim \mathcal{N}(0,E+N)$.

Hence, to optimally exploit the AWGN channel, there should be used NPM with Gaussian probability distribution. Instead, in applications there is used UPM, what as comes with a penalty.\\

Figure \ref{mutinf} presents mutual information and efficiency for two noise levels: $N=0.1$ and $N=1$. Surprisingly, for $N=1$, ternary and hexagonal modulation, throughput is maximized for $p=0$, what means not using the zero-signal. It is caused by the fact that zero-signal is nearly useless for such high level of noise - will be most likely interpreted as a different signal. However, the energy efficiency is maximized at the opposite end: for $p=1$, where for all modulations the limit is
\be \lim_{p\to 1} \eta = \frac{1}{N \ln(4)}\approx \frac{0.72135}{N} \ee

The $N=0.1$ case is more realistic. For example for hexagonal modulation, and UPM ($p=1/7$), one can transmit $\approx 2.568$ bits/energy unit. This amount can be doubled by using $p\approx 0.916$ frequency of zero signal, at cost of $\approx 5.1$ times lower mutual information (throughput). Finally $\lim_{p\to 1} \eta \approx 7.2135$, however, with throughput also going to 0.

\section{Reversed entropy coding + channel coding}
The standard entropy coding allows to encode a sequence of symbols using on average approximately Shannon entropy bits per symbol. It translates a sequence of symbols having some assumed probability distribution $\{p_s\}_s$ into a sequence of preferably uncorrelated $\Pr(0)=\Pr(1)=1/2$ bits (to maximize their informational content). To optimally use the assumed probability distribution, symbol of probability $p$ should on average use $\lg(1/p)$ bits of information, which generally does not have to be a natural number. For example $a\to 0,\ b\to 10,\ c\to 11$ is a prefix code optimal for $\Pr(a)=1/2,\ \Pr(b)=\Pr(c)=1/4$ probability distribution.

NPM requires to handle a kind of reversed entropy coding problem: the message to encode is a sequence of bits (preferably uncorrelated $\Pr(0)=\Pr(1)=1/2$ to maximize their content), and we want to translate it into a sequence of symbols of some chosen probability distribution. This purpose can be fulfilled by using entropy coder, but with switched encoder and decoder. For example a prefix code $0\to a,\ 10\to b,\ 11\to c$ can translate an i.i.d. $\Pr(0)=\Pr(1)=1/2$ input sequence of bits into a sequence of symbols with $\Pr(a)=1/2,\ \Pr(b)=1/4,\ \Pr(c)=1/4$ probability distribution.

However, this approach approximates probabilities with powers of 1/2 and cannot handle probability close to 1, useful for example for the discussed zero-signal. Additionally, adding error correction capabilities would require using some additional coding layer. We will discuss using recent tANS coding instead~(\cite{ANS,ANS1}), which has similar processing cost as Huffman coding (finite state automaton, no multiplication), but uses nearly accurate probabilities, including close to $1$. Additionally, it has a history dependent internal state, which can be used for error correction purposes in analogy to convolutional codes.

\subsection{Reversed tANS coding (rtANS)}
\begin{figure}[t!]
    \centering
        \includegraphics[width=8cm]{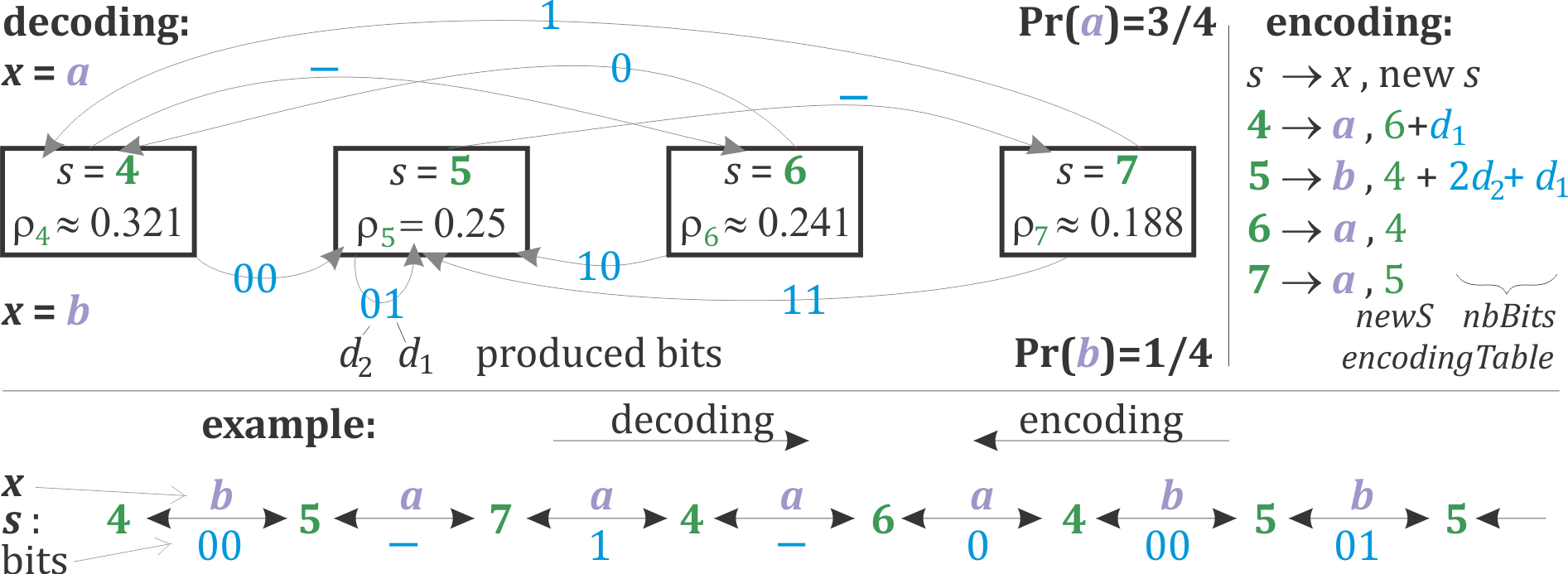}\
\begin{center}
        \caption{Example of reversed (encoder and decoder are switched) tANS automaton for $L=4$ states and $\Pr(a)=3/4,\ \Pr(b)=1/4$ probability distribution and its application for a bitstream (bottom). It was generated using $"abaa"$ symbol spread, with $L_a=3,\ L_b=1$ numbers of appearances, corresponding to the assumed probability distribution. Symbol $b$ carries $-\lg(1/4)=2$ bits of information and so the automaton always uses 2 bits for this symbol. In contrast, symbol $a$ carries $\lg(4/3)\approx 0.415$ bits of information. Hence the automaton sometimes uses one bit for this symbol, sometime 0 bits - just accumulating information in the state $s$. The $\rho_s$ denotes probability of using state $s$ by this automaton. Observe that decoding and encoding are performed in opposite directions. }
   \label{autom}
\end{center}
\end{figure}

Reversing tANS entropy coding, we constructs automaton with $L=2^R$ states, which translates a bit sequence into a sequence of symbols having a chosen probability distribution. For the purpose of this paper (it is usually opposite), we will denote state by $s\in\{L,\ldots,2L-1\}$ and symbol as $x\in\mathcal{A}$. The state $s$ acts as a buffer containing $\lg(s)\in [R,R+1)$ fractional number of bits. Symbol $x$ of probability $1/2^{k_x} \leq p_x < 1/2^{k_x-1}$ modifies the state and produces $k_x$ or $k_x-1$ bits to the bitstream, depending if $s$ is above or below some boundary. $L=4$ state example is presented in Fig. \ref{autom}. In practice there is for example used $L=2048$ states and $|\mathcal{A}|=256$ size alphabet.

The construction of such automaton first chooses quantized probability distribution: $L_x \in \mathbb{N}:\ \sum_x L_x = L$ and $p_x \approx L_x/L$. Then spread symbols: choose $symbol[s]$ for every position $s\in\{L,\ldots,2L-1\}$, such that symbol $x$ was used $L_x$ times.

The explanation and details can be found in other sources, here we focus on using for NPM. The reversed tANS (rtANS) encoding (original decoding) is presented as Algorithm \ref{dec0}. Each state determines symbol to produce, and a rule for new state: some base value ($newS$) and the number of bits to read from the bitstream ($nbBits$):

\begin{algorithm}[htbp]
\footnotesize{
\caption{rtANS encoding step, $S=s-L\in\{0,\ldots,L-1\}$}
\label{dec0}
\begin{algorithmic}
\STATE $t = encodingTable[S]$  \quad  \COMMENT{ $S\in \{0,..,L-1\}$ is current state }
\STATE useSymbol($t.symbol$)    \qquad\qquad   \COMMENT{ use or store decoded symbol }
\STATE $s = t.newS + $readBits$(t.nbBits)$   \qquad\qquad  \COMMENT{ state transition }
\end{algorithmic}
}
\end{algorithm}

This inexpensive loop would be used by the sender to transform bitstream (message) into stream of symbols of chosen probability distribution. Without getting into details here, the Algorithm \ref{gen} allows to generate the $encodingTable$, what can be done outside the device, and just written in it. This generation requires choosing $symbol[]$ spread function earlier, which determines the coding (has to be the same for decoding and encoding). Sources and benchmarks of various ways of symbol spread it can be found in~\cite{toolkit}.

\begin{algorithm}[htbp]
\footnotesize{
\caption{Preparation for rtANS encoding, $L=2^R$}
\label{gen}
\begin{algorithmic}
\REQUIRE $next[x] = L_x$ \quad\COMMENT{next appearance of symbol $x$}
\FOR {$S=0$ to $L-1$}
\STATE $t.symbol=symbol[S]$ \quad\COMMENT{ symbol is from spread function }
\STATE $s=next[t.symbol]++$
\STATE $t.nbBits = R -\lfloor \lg(s)\rfloor$ \qquad \COMMENT{ number of bits}
\STATE $t.newS = (s << t.nbBits)-L$\qquad\COMMENT{ properly shift $s$ }
\STATE $encodingTable[S]=t$
\ENDFOR
\end{algorithmic}
}
\end{algorithm}

Now let us look at the rtANS decoding step, which will be performed by receiver and is crucial from the point of view of eventual error correction. It is performed in opposite direction then encoding: it starts with the final encoding step (needs to be transmitted) and ends with the initial encoding state (arbitrary, may contain some information). For the discussed purpose encoding can be made in forward direction, decoding in backward. For simplicity we will discuss decoding as in forward direction here.

As in Figure \ref{autom}, we can write decoding step as
\be D(s,x)=(s_{sx},B_{sx}) \label{decoding} \ee
where $s$ is the starting state. Symbol $x$ takes us to state $s_{sx}$, and produces $B_{sx}\in\mathcal{A}^*$ bit sequence of length $|B_{sx}|\in\{0,\ldots,\lg(L)\}$ (can be empty). Additionally, $B_{xs}$ are just $|B_{xs}|\approx \lg(1/p_x)$ youngest bits of the state $s$. Optimized implementation and preparation is presented as Algorithms \ref{enc0} and \ref{encprep}:

\begin{spacing}{0.5}
\begin{algorithm}[htbp]
\footnotesize{
\caption{rtANS decoding step for symbol $x$ and state $s=S+L$}
\label{enc0}
\begin{algorithmic}
\STATE $nbBits = (s + nb[x]) >> r$    \quad\qquad \COMMENT{$r=R+1,\ 2^r = 2L$}
\STATE useBits$(s, nbBits)$  \qquad\COMMENT {use $nbBits$ of the youngest bits of $s$}
\STATE $s = decodingTable[start[x] + (s >> nbBits)]$
\end{algorithmic}
}
\end{algorithm}
\end{spacing}

\begin{algorithm}[htbp]
\footnotesize{
\caption{Preparation for rtANS decoding, $L=2^R$, $r=R+1$}
\label{encprep}
\begin{algorithmic}
\REQUIRE $k[x] = R-\lfloor \lg(L_x) \rfloor$  \qquad \COMMENT{$nbBits=k[x]$ or $k[x]-1$}
\REQUIRE $nb[x] = (k[x] << r)-(L_x << k[x])$
\REQUIRE $start[x]= - L_x + \sum_{i < x} L_i$
\REQUIRE $next[x] = L_x$
\FOR {$s=L$ to $2L-1$}
\STATE $x=symbol[s-L]$
\STATE $decodingTable[start[x] + (next[x]++)] = s$;
\ENDFOR
\end{algorithmic}
}
\end{algorithm}

\subsection{Forward error correction}
We will now discuss adding forward error correction: sender adds redundancy in the transmitted data (inexpensive), which is used by receiver to correct eventual errors (relatively expensive). A simple way to add redundancy for rtANS is for example by putting the message in even positions of the bitstream to encode, and 0 in odd positions, getting rate 1/2 code. For simplicity we will focus on this case, but analogously we can get different rational rates, for example 2/3 by putting 0 every 3rd position.

Let us denote $U=\{u_0,\ldots, u_{N-1}\}$ as encoded bitstream, which even positions contain the message, and odd positions are zeros: $\forall_i\  u_{2i+1}=0$. The rtANS has transformed it into the transmitted sequence of symbols $\{x_0,\ldots, x_{n-1}\}$, which was modified by the noise into received $\{y_0,\ldots,y_{n-1}\}$ sequence of symbols. The goal of receiver is to find the closest sequence $X$, for which decoded bitstream has zeros at odd positions.

Let us denote $\{P_0=0,\ P_1 ,\ldots, P_n =N\}$ as positions in $U$ corresponding to successive symbols, and $\{s_0,\ldots,s_n\}$ as states of encoder, such that:
\be D(s_t,x_t)=(s_{t+1}, \{U_{P_t},\ U_{P_t+1},\ldots, U_{P_{t+1}-1}\}) \ee
It assumes for simplicity forward decoding. State $s_0$ is initial for decoding (final for encoding, also transmitted), $s_n$ is the final state of decoding (initial for encoding).\\

Error correction in our case has similar inconvenience as for synchronization channels: the relation between symbols and corresponding bit blocks of varying length ($P_t$ sequence) is not known. It is a crucial complication for Viterbi~\cite{viterbi} and BCJR~\cite{bcjr} type of approaches. Fortunately, this is not an issue for sequential decoding~\cite{fano}, also successfully applied for deletion channel~\cite{deletion}.

In sequential decoding we build a tree of corrections: start with the root as known state $s_0$ and expand it, such that each branch in depth $t$ corresponds to one of possible choices of $x_t$. As expanding all possibilities would mean exponential growth, it is crucial to expand only promising looking nodes. In this case, the number of nodes to consider for reasonable parameters is usually a relatively small multiplicity of the length of sequence - tools for analysis can be found in \cite{cortree}.

A depth $t$ node of the tree corresponds to assuming some sequence of symbols $\{x_0,\ldots,x_{t-1}\}$, which corresponds to some $\{u_0,\ldots, u_{T-1}\}$ hypothetical prefix of the bit sequence. The applied redundancy says to consider only nodes fulfilling $\forall_i\ u_{2i+1}=0$, other branches are not expanded. Length $T$ bit sequence has $\lfloor T/2 \rfloor$ such bits verifying that we consider a prefix of a codeword. For an improper correction (node) we can assume that these bits are i.i.d. $\Pr(0)=\Pr(1)=1/2$, so there is $2^{-\lfloor T/2 \rfloor}$ probability of accidentally fulfilling these constraints. While choosing a leaf to expand, Bayes analysis says that probability that a given leaf is the proper one is proportional to:

$$\frac{\Pr(\{x_0,\ldots,x_{t-1}\}|\{y_0,\ldots,y_{t-1}\})}{\Pr(\textrm{accidentially fulfilling the constraints})}\propto$$
\be 2^{\lfloor T/2 \rfloor} \prod_{0\leq i < t} (\Pr(x_i) \Pr(y_i|x_i)) \ee

\noindent where $\Pr(x_i)$ is the chosen probability distribution of symbols, $\Pr(y_i|x_i)$ is the assumed model of error, for example Gaussian distribution. In practice there is used weight $w$ as logarithm of the above formula, calculated for new node as weight of its father plus $\Delta w$:
\be \Delta w = \lg(\Pr(x_t) \Pr(y_t|x_t)) + \#\{\textrm{new constrained bits}\} \ee
with the number of positions constrained by the rule $u_{2i+1}=0$ in the new bit block corresponding to using symbol $x_t$.

Finally, the correction process after initialization with a root before the first symbol, is a loop of choosing the most promising leaf so far (having largest $w$) and expanding it, until reaching the final position with the proper final state.

The rtANS decoding has convenient property that bits produced in given step are some number of the least significant bits of the state $s$. Knowing the state we can quickly determine the maximal number of such bits to fulfill the $b_{2i+1}=0$ condition, which determines the symbols which could be used in this step.

\section{Conclusions}
While the standard approach is encoding information using uniform probability distribution among some symbols (UPM), we have discussed practical application of non-uniform distributions (NPM). For example capacity of AWGN channel is fulfilled for Gaussian distribution, not the uniform one. Instead of prioritizing on channel capacity, we have focused on energy efficiently here: amount of transmitted bits per energy unit, which can be increased at cost of reduced throughput. It can be practically doubled for hexagonal modulation, or quadrupled for binary modulation, by more frequent use of zero-signal. Example of application is improving battery life of remote sensors.

The discussed solution for coding was tANS entropy coder, which has inexpensive processing cost (finite state automaton, no multiplication), but uses nearly accurate probabilities.  Additionally, there was discussed cost-free redundancy addition while this encoding step, for example by inserting zeros at odd positions of the bit sequence. Sequential decoding can be used for error correction of such message. It is slightly more complex than for UPM, but this energy and hardware cost is not paid in the remote sensor.
\bibliographystyle{IEEEtran}
\bibliography{ref}

\begin{thebibliography}{10}
\providecommand{\url}[1]{#1}
\csname url@samestyle\endcsname
\providecommand{\newblock}{\relax}
\providecommand{\bibinfo}[2]{#2}
\providecommand{\BIBentrySTDinterwordspacing}{\spaceskip=0pt\relax}
\providecommand{\BIBentryALTinterwordstretchfactor}{4}
\providecommand{\BIBentryALTinterwordspacing}{\spaceskip=\fontdimen2\font plus
\BIBentryALTinterwordstretchfactor\fontdimen3\font minus
  \fontdimen4\font\relax}
\providecommand{\BIBforeignlanguage}[2]{{%
\expandafter\ifx\csname l@#1\endcsname\relax
\typeout{** WARNING: IEEEtran.bst: No hyphenation pattern has been}%
\typeout{** loaded for the language `#1'. Using the pattern for}%
\typeout{** the default language instead.}%
\else
\language=\csname l@#1\endcsname
\fi
#2}}
\providecommand{\BIBdecl}{\relax}
\BIBdecl

\bibitem{d2007}
J.~Duda, ``Optimal encoding on discrete lattice with translational invariant
  constrains using statistical algorithms,'' \emph{arXiv preprint
  arXiv:0710.3861}, 2007.

\bibitem{KT1}
------, ``Embedding grayscale halftone pictures in qr codes using correction
  trees,'' \emph{arXiv preprint arXiv:1211.1572}, 2012.

\bibitem{KT}
J.~Duda, P.~Korus, N.~J. Gadgil, K.~Tahboub, and E.~J. Delp, ``Image-like 2d
  barcodes using generalizations of the kuznetsov--tsybakov problem,''
  \emph{IEEE Transactions on Information Forensics and Security}, vol.~11,
  no.~4, pp. 691--703, 2016.

\bibitem{ANS}
J.~Duda, ``Asymmetric numeral systems: entropy coding combining speed of
  huffman coding with compression rate of arithmetic coding,'' \emph{arXiv
  preprint arXiv:1311.2540}, 2013.

\bibitem{ANS1}
J.~Duda, K.~Tahboub, N.~J. Gadgil, and E.~J. Delp, ``The use of asymmetric
  numeral systems as an accurate replacement for huffman coding,'' in
  \emph{Picture Coding Symposium (PCS), 2015}.\hskip 1em plus 0.5em minus
  0.4em\relax IEEE, 2015, pp. 65--69.

\bibitem{conv}
P.~Elias, ``Coding for noisy channels,'' \emph{IRE Conv. Rec. Part 4}, 1955.

\bibitem{shannon}
C.~E. Shannon, ``A mathematical theory of communication,'' \emph{The Bell
  System Technical Journal}, vol.~27, no.~3, pp. 379--423, 1948.

\bibitem{toolkit}
{J. Duda}, {https://github.com/JarekDuda/AsymmetricNumeralSystemsToolkit}.

\bibitem{viterbi}
A.~Viterbi, ``Error bounds for convolutional codes and an asymptotically
  optimum decoding algorithm,'' \emph{IEEE transactions on Information Theory},
  vol.~13, no.~2, pp. 260--269, 1967.

\bibitem{bcjr}
L.~Bahl, J.~Cocke, F.~Jelinek, and J.~Raviv, ``Optimal decoding of linear codes
  for minimizing symbol error rate,'' \emph{IEEE Transactions on Information
  Theory}, vol.~20, no.~2, 1974.

\bibitem{fano}
R.~Fano, ``A heuristic discussion of probabilistic decoding,'' \emph{IEEE
  Transactions on Information Theory}, vol.~9, no.~2, pp. 64--74, 1963.

\bibitem{deletion}
{J. Duda}, {https://github.com/JarekDuda/DeletionChannelPracticalCorrection}.

\bibitem{cortree}
J.~Duda and P.~Korus, ``Correction trees as an alternative to turbo codes and
  low density parity check codes,'' \emph{arXiv preprint arXiv:1204.5317},
  2012.

\end{thebibliography}
\end{document}